\documentclass[journal=jcisd8,manuscript=article]{achemso}
\SectionNumbersOn

\usepackage{microtype}
\usepackage{graphicx}
\usepackage{float}
\usepackage{subcaption}
\usepackage{booktabs}
\usepackage{amsmath}
\usepackage{amssymb}
\usepackage{mathtools}
\usepackage{amsthm}
\usepackage{url}
\usepackage{enumitem}
\usepackage{tabularx}
\usepackage{siunitx}
\usepackage{xcolor}
\usepackage[most]{tcolorbox}
\usepackage{caption}
\usepackage{verbatim}
\usepackage{hyperref}
\usepackage[capitalize,noabbrev]{cleveref}

\newif\ifrevision
\revisionfalse
\providecommand{\textalpha}{\ensuremath{\alpha}}

\ifrevision
  \newcommand{\new}[1]{\textcolor{blue}{#1}}
  \newcommand{\newtable}{\color{blue}}
\else
  \newcommand{\new}[1]{#1}
  \newcommand{\newtable}{}
\fi

\newtcolorbox{userbox}{
  colback=blue!5!white,
  colframe=blue!20!white,
  boxsep=5pt,
  arc=4mm,
  left=10pt,
  right=10pt,
  top=10pt,
  bottom=10pt,
  breakable,
}

\newtcolorbox{assistantbox}{
  colback=black!5!white,
  colframe=black!20!white,
  boxsep=5pt,
  arc=4mm,
  left=10pt,
  right=10pt,
  top=10pt,
  bottom=10pt,
  breakable,
}
\author{Carles Navarro}
\email{c.navarro@acellera.com}
\affiliation[Acellera Labs]{Acellera Labs, C/ Doctor Trueta 183, 08005 Barcelona, Spain}
\alsoaffiliation[Universitat Pompeu Fabra]{Computational Science Laboratory, Universitat Pompeu Fabra, PRBB, C/ Doctor Aiguader 88, 08003 Barcelona, Spain}

\author{Mariona Torrens-Fontanals}
\affiliation[Acellera Labs]{Acellera Labs, C/ Doctor Trueta 183, 08005 Barcelona, Spain}

\author{Philipp Tholke}
\affiliation[Acellera Labs]{Acellera Labs, C/ Doctor Trueta 183, 08005 Barcelona, Spain}

\author{Stefan Doerr}
\affiliation[Acellera Labs]{Acellera Labs, C/ Doctor Trueta 183, 08005 Barcelona, Spain}

\author{Gianni de Fabritiis}
\email{g.defabritiis@acellera.com}
\affiliation[Acellera Labs]{Acellera Labs, C/ Doctor Trueta 183, 08005 Barcelona, Spain}
\alsoaffiliation[Universitat Pompeu Fabra]{Computational Science Laboratory, Universitat Pompeu Fabra, PRBB, C/ Doctor Aiguader 88, 08003 Barcelona, Spain}
\alsoaffiliation[ICREA]{Instituci\'{o} Catalana de Recerca i Estudis Avan\c{c}ats (ICREA), Passeig Llu\'{i}s Companys 23, 08010 Barcelona, Spain}

\title[Speak to a Protein: An Interactive Multimodal Co-Scientist]{Speak to a Protein: An Interactive Multimodal Co-Scientist for Protein Analysis}
\keywords{Agentic co-scientist, scientific discovery, molecular visualization, deep research, drug discovery, retrieval-augmented generation}

\begin{document}

\begin{tocentry}
\centering
\includegraphics[width=3.25in]{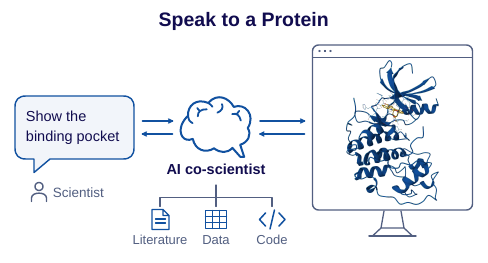}
\end{tocentry}

\begin{abstract}
Building a working mental model of a protein typically requires weeks of reading, cross-referencing crystal and predicted structures, and inspecting ligand complexes, an effort that is slow, unevenly accessible, and often requires specialized computational skills. We introduce \emph{Speak to a Protein}, a new capability that turns protein analysis into an interactive, multimodal dialogue with an expert co-scientist. The AI system retrieves and synthesizes relevant literature, structures, and ligand data; grounds answers in a live 3D scene; and can highlight, annotate, manipulate and see the visualization. It also generates and runs code when needed, explaining results in both text and graphics.
We demonstrate these capabilities on relevant proteins, posing questions about binding pockets, conformational changes, or structure-activity relationships to test ideas in real time. \emph{Speak to a Protein} reduces the time from question to evidence, lowers the barrier to advanced structural analysis, and enables hypothesis generation by tightly coupling language, code, and 3D structures. 
\emph{Speak to a Protein} is freely accessible at \url{https://open.playmolecule.org}.

\end{abstract}

\section{Introduction}\label{sec:intro}
Proteins are the molecular machinery of life, and understanding their structure and function is fundamental to modern biology and medicine. For a researcher in drug discovery or molecular biology, developing an intuitive, ``working mental model'' of a target protein, its active sites, its conformational dynamics, and its network of interactions is a critical first step. However, this process is slow and arduous, often requiring a combination of deep domain knowledge and specialized computational skills.

A researcher investigating a protein kinase to understand how a new series of inhibitors might bind must embark on a fragmented and technically demanding workflow. This involves sifting through the PubMed literature~\citep{Sayers2020}, fetching and comparing multiple structures from the Protein Data Bank (PDB)~\citep{Burley2018}, querying UniProt~\citep{UniProt2022}  for functional annotations and disease-associated variants, and extracting structure-activity relationship (SAR) data from databases like ChEMBL~\citep{Mendez2018}. Each step requires navigating different interfaces and data formats. Furthermore, deeper analysis, such as superimposing structures, identifying key interactions, or plotting bioactivity data, requires proficiency with specialized software or scripting languages, creating a significant barrier to entry for many bench scientists. This high friction for asking and answering questions restrains curiosity and slows the pace of discovery.

To address these challenges, we introduce \emph{Speak to a Protein}, a new capability designed to transform protein analysis into an interactive dialogue with an AI co-scientist that collaborates with the user in real time. Using recent advances in large language models (LLMs) and their evolution into multimodal scientific agents \citep{Caffagni2024, Zhang2024, Zheng2025}, our system can comprehend complex, natural language queries about a protein of interest. It autonomously retrieves, integrates, and synthesizes information from a comprehensive suite of biological data sources, including literature, structural repositories, and biochemical databases.

The core innovation of \emph{Speak to a Protein} is its ability to ground its responses across multiple, synchronized modalities. When a user asks a question, the AI does not simply return text. It interacts with a live 3D structural viewer to highlight residues, measure distances, or annotate binding pockets. It can generate and execute code in a sandboxed environment to perform calculations, filter tabular data, or generate plots on the fly. Furthermore, the AI co-scientist sees, understands and controls the visualization, offering a natural interaction with the user. This tight coupling of natural language, 3D visualization, and code execution creates a seamless and intuitive environment for scientific exploration.

This paper makes the following contributions: We present the design and architecture of an AI system that integrates language, code execution, and 3D visualization for interactive protein analysis. We show how this multimodal approach drastically lowers the barrier to complex structural and biochemical data analysis. Through case studies on relevant proteins and benchmarks, we show that this system enables a more fluid and powerful form of hypothesis generation, accelerating the cycle from question to evidence.

\section{Related Work}\label{sec:related}
The ambition to create an AI capable of scientific discovery is a long-standing goal, articulated in visions such as Hiroaki Kitano's proposal for an ``AI Scientist'': a system that could autonomously formulate hypotheses, design experiments, and achieve Nobel-class discoveries \citep{kitano2021}. Recent systems push this frontier further: \emph{Kosmos} \citep{mitchener2025kosmos} runs multi-hour autonomous campaigns that read thousands of papers and execute tens of thousands of lines of analysis code per task. While such fully autonomous systems \citep{boiko2023autonomous, zou2025agente, mitchener2025kosmos} remain a grand challenge, recent progress in large language models (LLMs) has enabled the development of a more immediate and collaborative paradigm: the ``AI co-scientist'' \citep{gottweis2025coscientist} or ``advanced intelligence'' in Kitano's words.

Early systems explored conversational interfaces for structural inspection and Q\&A over proteins. \citet{guo2023proteinchat} demonstrated \emph{ProteinChat}, which couples LLM prompting with protein 3D structures to answer user questions about residues and pockets. Contemporary efforts such as \citet{wang2024protchatgpt} and \citet{xiao2024proteingpt} investigate protein-aware prompting and multimodal conditioning for function/property reasoning. Most recently, \citet{wang2025prot2chat} proposes \emph{Prot2Chat}, an LLM that fuses protein sequence, structure, and text via an early-fusion adapter, directly targeting protein Q\&A. Complementary work strengthens the protein side itself, with structure-enhanced instruction tuning \citep{wu2025sepit} and trimodal alignment of sequence, structure, and language for advanced protein search \citep{su2025protrek}.

Beyond text-only chat, domain copilots increasingly drive molecular viewers and modeling tools. \citet{sun2024chatmol} introduce \emph{ChatMol Copilot}, an LLM agent that coordinates cheminformatics and modeling tools (e.g., docking, conformer generation) in response to natural-language requests. In parallel, \citet{ille2024gpt4structural} systematically evaluates GPT-4's ability to perform rudimentary structural modeling and protein–ligand interaction analysis, highlighting both promise and limitations. Our work is aligned with this line but centers on tightly coupling language, code execution, and a live 3D scene for grounded, interactive answers.

Agent frameworks augment LLMs with tool use, retrieval, and planning. \emph{ChemCrow} \citep{bran2024chemcrow} shows that equipping GPT-4 with chemistry tools enables multi-step synthesis planning and materials tasks. More recently, CLADD \citep{lee2025cladd} proposes a retrieval-augmented multi-agent system specialized for drug discovery tasks. Multi-agent platforms extend this paradigm to end-to-end workflows: \emph{LLM-RDF} \citep{ruan2024llmrdf} coordinates six specialized agents across literature scouting, experiment design, hardware execution, and result interpretation, while \emph{ChatInvent} \citep{he2026chatinvent} brings agentic molecular design and synthesis planning into a real-world pharmaceutical setting. At broader scope, \emph{Biomni} \citep{huang2025biomni} pairs LLM reasoning with retrieval-augmented planning and code execution over a large catalogue of biomedical tools and databases, targeting general-purpose biomedical workflows. \emph{Speak to a Protein} adopts the agentic paradigm for structural biology: it retrieves literature/structures, executes analyses, and grounds responses in synchronized 3D visualizations.

\section{System Overview: \emph{Speak to a Protein}}\label{sec:overview}

\subsection{Architecture}

\begin{figure}[H]
\centering
\includegraphics[width=\textwidth]{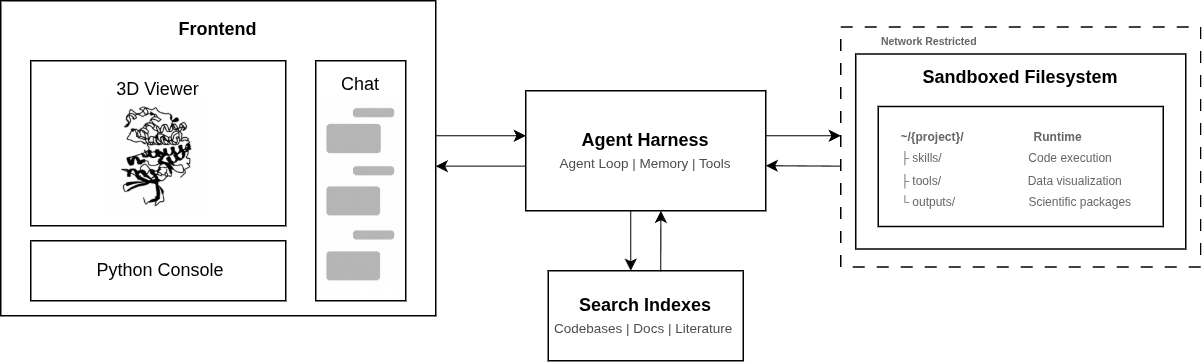}
\caption{\textbf{Overview of the system architecture.} The Frontend hosts a chat panel, a 3D molecular viewer, and a Python console. The Agent Harness runs the agent loop, holds conversation memory, and executes tool calls. A search service indexes codebases, documentation, and literature. A sandboxed filesystem provides a per-project workspace and a preinstalled scientific software stack, with restricted network egress.}\label{fig:architecture}
\end{figure}

The system architecture of \emph{Speak to a Protein} (\cref{fig:architecture}) is organized around four components: a frontend with a chat panel and a 3D viewer where the user and the agent inspect and manipulate structures for 3D reasoning, an agent harness that runs the agent loop and exposes tools to the language model, a search service that maintains semantic indices over codebases, documentation, and literature, and a sandboxed filesystem in which the agent runs code against a preinstalled scientific software stack.

The frontend hosts the chat panel, the 3D molecular viewer (\cref{fig:viewer} and \cref{subsec:viewer}), and a Pyodide-based Python console for client-side analysis. The user converses through the chat panel; the frontend renders the agent's text replies and applies viewer commands as the agent issues them.

At each user request, the agent harness drives an agent loop: the model emits tool calls, the harness dispatches them, and their outputs are fed back into the next model invocation, continuing until the task is complete. To handle long sessions, the memory module autocompacts older turns into a working memory: a compact, adaptive summary that the agent itself rewrites at each compaction and that is reinjected into context on subsequent turns.

\subsection{Tools}\label{sec:tools}

The system exposes several tools to the agent, a set of core primitives that run code and edit files inside the sandbox, viewer tools that drive the 3D viewer in the frontend, a semantic search service over indexed library codebases, and a literature retrieval tool that gathers relevant publications for a given protein. \new{Each tool is registered in the harness and presented to the model as a function call; \cref{tab:tools} lists the complete set available in the deployment described here.}

\noindent\textbf{Core tools.} The system's core tools are a set of primitives that execute shell commands in the sandbox, read, write, edit, and visualize files there, and delegate self-contained subtasks to sub-agents. All execution happens inside the sandboxed filesystem of \cref{fig:architecture}, a Docker container with a \new{per-project} workspace, a preinstalled scientific software stack, and outbound network restricted by a forward proxy that allowlists scientific domains. The stack spans Python, R with Bioconductor, and command-line bioinformatics tools. Files persist across turns, supporting iterative refinement and reproducibility.

\noindent\textbf{Viewer Tools.} The viewer tools provide the functionality to load project files into the viewer, execute code in the viewer's Python session for visualization control and structural manipulation, capture screenshots of the rendered state, and save modified objects back to the project workspace. The bridge between the harness and the viewer is a session-bound WebSocket, each chat holds a single persistent channel. Screenshots are returned to the agent automatically after most viewer operations, providing a vision channel for grounding subsequent reasoning. \new{The viewer runs its own Python session in the browser, separate from the sandbox, and the two share no state. Which tool the agent calls therefore determines where the code runs, in the browser session or in the container, and files are the only bridge between the two, moved by the load and save tools.}

\noindent\textbf{Codebase Search.} The codebase search service holds vector indices over library codebases and their documentation, including MoleculeKit \citep{Doerr2016} and HTMD \citep{Doerr2016}. This provides the agent with a fast semantic search over the codebase to inspect and discover how to use these scientific packages.

\noindent\textbf{Literature Search.}\label{sec:data:literature} The literature tool assembles a protein-conditioned RAG corpus on demand. While the corpus is built from \emph{curated references from UniProt}, it accepts a UniProt accession, a PDB identifier, or a FASTA sequence; each is resolved to a set of UniProt entries (a PDB code maps through its cross-references, a FASTA sequence is matched against PDB structures and then to UniProt), and the references listed under those entries form the corpus. Open-access articles are fetched from PubMed Central and indexed with LlamaIndex~\citep{Liu_LlamaIndex_2022} into a per-protein vector store cached on disk and keyed on the FASTA hash (Table~S1). At query time the agent issues a free-text query and receives the top-$k$ relevant passages with their citations. Index construction completes in one to two minutes for a new protein; subsequent queries on the same target are served from cache in seconds.

\begin{table}[H]
\caption{\new{\textbf{Tools available to the agent.} The agent calls each tool by name. Sandbox and file tools act on the per-project workspace; viewer tools act on the 3D scene in the frontend through the session WebSocket.}}\label{tab:tools}
{\newtable
\small
\begin{tabularx}{\textwidth}{llX}
\toprule
\textbf{Group} & \textbf{Tool} & \textbf{Function} \\
\midrule
Sandbox & \texttt{run\_bash} & Runs a bash command in the container. \\
        & \texttt{cancel\_task} & Stops a background command. \\
\addlinespace
Files   & \texttt{read\_file} & Reads a text file. \\
        & \texttt{write\_file} & Writes a file. \\
        & \texttt{edit\_file} & Replaces text inside a file. \\
        & \texttt{view\_image} & Returns an image to the agent. \\
\addlinespace
Viewer  & \texttt{pmview\_load} & Loads workspace files into the viewer. \\
        & \texttt{pmview\_run\_python} & Runs Python in the viewer session. \\
        & \texttt{pmview\_screenshot} & Saves an image of the viewer state. \\
        & \texttt{pmview\_save} & Saves a viewer object to the workspace. \\
\addlinespace
Search  & \texttt{codebase\_search} & Searches indexed codebases and documentation. \\
        & \texttt{query\_protein\_literature} & Retrieves passages from the literature corpus. \\
\addlinespace
Delegation & \texttt{run\_agent} & Starts a sub-agent on a subtask. \\
        & \texttt{message\_agent} & Sends a message to a sub-agent. \\
\bottomrule
\end{tabularx}
}
\end{table}

\subsection{Scientific Skills}\label{sec:skills}

\emph{Skills} are documentation packages that teach the agent how to use a particular database, library, or workflow~\citep{anthropic_skills_2025}. Each skill describes the relevant inputs and identifiers, the APIs to call, and idiomatic usage patterns, and may bundle example scripts the agent can adapt. Skills are exposed through progressive disclosure: at session start the agent receives only a one-line summary of every available skill, and reads the full body when one becomes relevant.

We currently provide skills for protein and structure databases (UniProt, PDB, AlphaFoldDB~\citep{FLEMING2025168967}), chemical and bioactivity databases (ChEMBL, PubChem~\citep{10.1093/nar/gkae1059}, DrugBank~\citep{10.1093/nar/gkad976}, ZINC~\citep{Irwin2005}), gene and genomic databases (NCBI Gene~\citep{Sayers2024}, Ensembl~\citep{10.1093/nar/gkae1071}), and biomedical literature retrieval (PubMed). New databases or workflows are added by authoring a new skill, without modifying the core system.

\subsection{Agentic 3D Grounding and Interfaces with Scientists}\label{subsec:viewer}

Understanding proteins requires the navigation of multiple types of information, such as three-dimensional structures, experimental assay tables, and textual annotations. A central goal of \emph{Speak to a Protein} is to connect these diverse modalities through a single conversational interface, ensuring that system responses are not only generated in natural language but are also grounded in concrete evidence such as 3D visualizations, data tables, and executable code. This grounding operates bidirectionally: users can visually verify AI-generated hypotheses, while the AI agent itself can capture and interpret viewer states to inform its reasoning. To achieve this, we provide a set of interactive interfaces that extend dialogue into complementary domains: a structural viewer for molecular inspection, a tabular analysis environment for filtering and plotting assay data, and mechanisms for synchronizing actions across views. Together, these interfaces enable users to fluidly transition between asking questions, running analyses, and visually verifying hypotheses. Significantly, we are building these capabilities on top of a web application that has already attracted more than 21,000 registered users over the years, even in the absence of these AI features. \new{This number counts every user registered on the platform since its launch, while the AI chat opened in October 2025 and has been used by 1,921 of them.} \new{Because the analysis is driven by dialogue, a user who does not write code can carry out work that would otherwise require scripting, while the code that produces each result stays visible in the chat.} We thus anticipate that it will be used by a considerable number of scientists, as demonstrated by the platform usage metrics in Appendix~S1.2.

\new{The intended user is a scientist who knows the biology of the target but does not necessarily write code. The system keeps the work checkable for both kinds of user, since the code the agent runs is shown in the chat and the viewer shows what was selected and measured. A user can also ask the agent to cite the source of every claim and to save the scripts it runs, and the session record keeps the code of every step, so the analysis can be repeated later.}

In \emph{Speak to a Protein}, these functionalities are built using an entirely client-side sandbox for dynamic visualization and manipulation of molecular structures \citep{torrens2023playmolecule}. The sandbox builds on a browser-based molecular visualization toolkit that combines the high-performance \texttt{mol*} visualization engine \citep{Sehnal2021},  capable of rendering large biomolecular structures and molecular dynamics trajectories directly in the browser, with a WebAssembly-enabled Python runtime (Pyodide). This allows the use of powerful Python libraries such as MoleculeKit \citep{Doerr2016} in the client environment, enhancing the viewer's capabilities to load and manipulate a wide range of common structural file formats, from PDB and CIF to molecular dynamics trajectory files such as XTC and TRR.

\begin{figure}[H]
\centering
\includegraphics[width=\columnwidth]{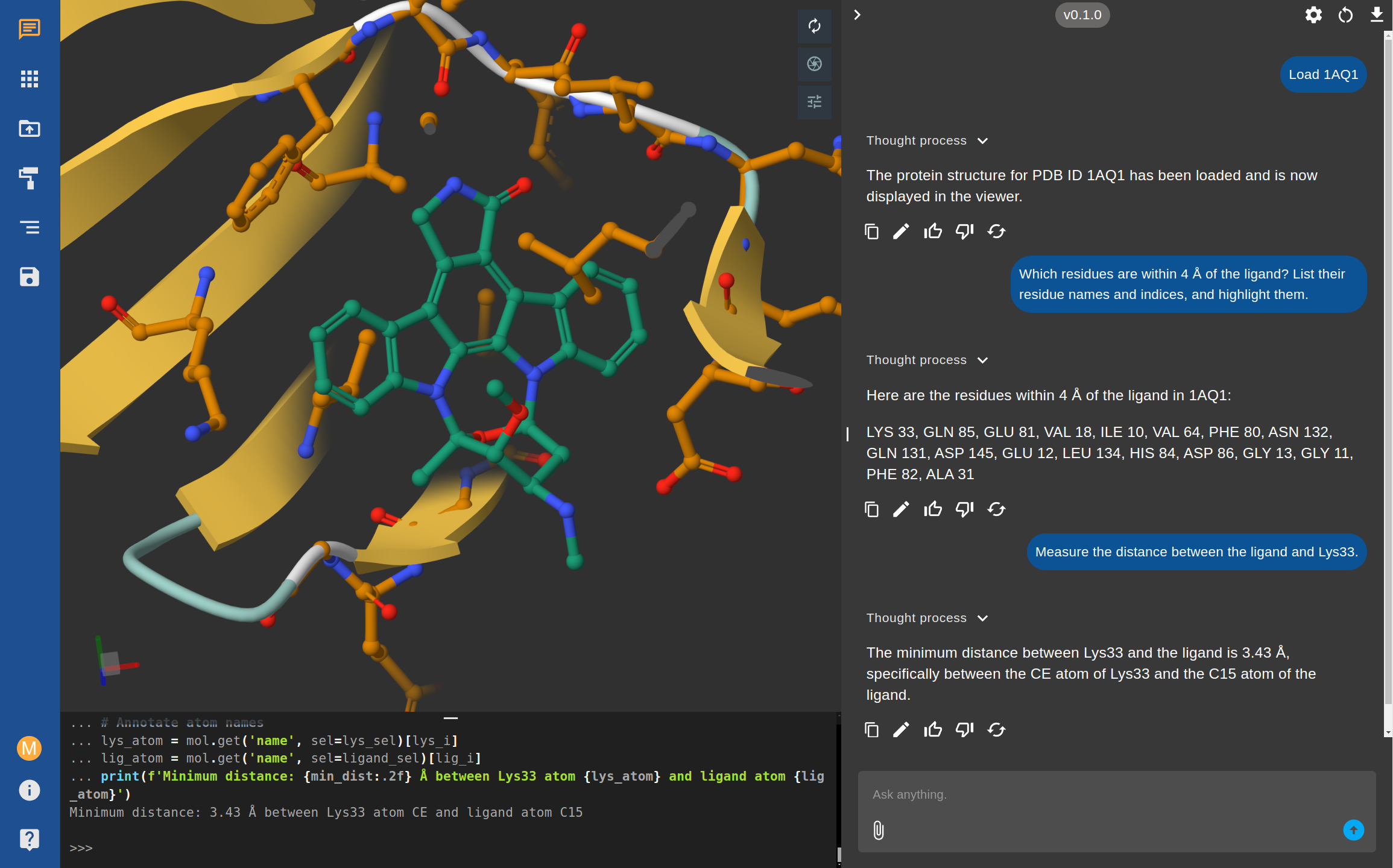}
\caption{
    \textbf{Interactive analysis of the CDK2 structure (PDB: 1AQ1) using the \emph{Speak to a Protein} multimodal assistant.}
    The user enters natural language queries in the chat panel (right), and the system responds with both textual answers and real-time updates to the 3D visualization (left). Python code executed in the \new{viewer's browser Python session} is shown in the console (bottom), providing full transparency and reproducibility of the underlying analyses.
}\label{fig:viewer}
\end{figure}

A key feature is the ability to control this viewer through natural language commands. Requests such as ``highlight the ATP-binding site'' are translated into tool calls that execute Python code within the viewer, producing visual changes in real time (\cref{fig:viewer}). Beyond user-initiated actions, the agent can autonomously capture screenshots of the current viewer state and analyze them using vision capabilities. This agentic 3D vision enables the model to verify the results of its own manipulations, identify structural features that may not be explicitly encoded in the data, and iteratively refine visualizations based on what it observes. \new{The agent does not track the contents of the viewer from the conversation history. Every user message carries the current viewer state with it, the loaded systems, tables and plots with their identifiers and visibility, a mark on any object the user has edited by hand, the current atom selection, and a screenshot of the scene, so the agent works from what the viewer reports and remains correct when the user changes the scene through the console or the interface.}

\section{Experiments}\label{sec:cases}
Evaluating agentic systems for computational biology is challenging: most existing benchmarks probe isolated capabilities rather than integrated, multi-step scientific work. We assess \emph{Speak to a Protein} on \emph{BixBench-Verified-50} \citep{phylo2026bixbenchverified} and complement the benchmark with two case studies on the dopamine D3 receptor (D3R) \citep{Chien2010} and cyclin-dependent kinase 2 (CDK2) \citep{Malumbres2009}, which illustrate how the system integrates data from multiple sources, generates insights through interactive multimodal analysis, and produces a structured LaTeX report that can be shared and queried later. All experiments use GPT-5.4 as the foundational language model driving the agent.

\subsection{BixBench-Verified-50}\label{subsec:bixbench}
\emph{BixBench} \citep{mitchener2025bixbench} comprises real-world computational biology scenarios with open-answer questions covering data exploration, multi-step analytical workflows, and interpretation of nuanced results. Frontier models on the original benchmark scored near 17\% in the open-answer regime, exposing both genuine agent limitations and substantial benchmark noise. \emph{BixBench-Verified-50} \citep{phylo2026bixbenchverified} is a 50-task curated subset in which domain experts removed or rewrote ambiguous items and corrected faulty ground truths, isolating actual agent capability. \new{This subset was released in a blog post and not in a peer-reviewed venue, which is a limitation of the comparison. We still prefer it to the original set, because the ambiguous items and the faulty ground truths that the curation removed would otherwise weigh on the score as much as the ability of the agent.}

\begin{figure}[H]
\centering
\includegraphics[width=0.85\columnwidth]{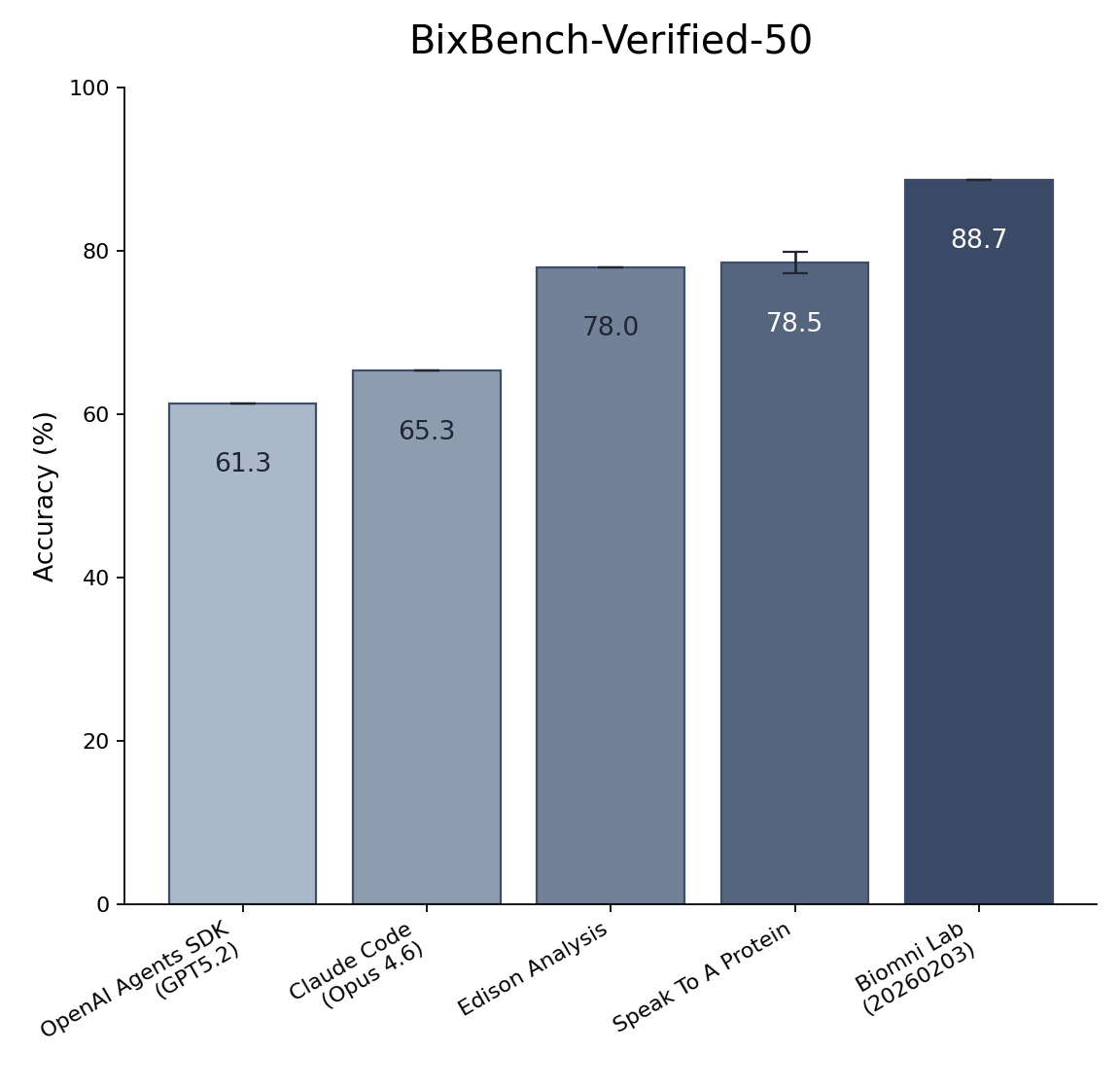}
\caption{Accuracy on BixBench-Verified-50. Baseline scores are reproduced from the public leaderboard; error bars for \emph{Speak to a Protein} indicate variability across repeated runs.}
\label{fig:bixbench}
\end{figure}

BixBench-Verified-50 emphasises bioinformatics analysis tasks (sequencing, single-cell, statistical genomics) that partially overlap with the structural biology workflows \emph{Speak to a Protein} primarily targets, motivating the complementary case studies that follow. On this benchmark \emph{Speak to a Protein} reached \textbf{78.5\%} accuracy, surpassing strong general-purpose coding agents such as Claude Code with Opus 4.6 (65.3\%) and the OpenAI Agents SDK with GPT-5.2 (61.3\%), and matching Edison Analysis at 78.0\% (\cref{fig:bixbench}). The leading score is held by Biomni Lab \citep{huang2025biomni} at 88.7\%, a system whose tool catalogue and curated retrieval corpus are explicitly tuned for the bioinformatics workflows that dominate this benchmark. We read the result as encouraging: \emph{Speak to a Protein}'s agentic substrate transfers well outside its structural biology focus, while the remaining gap to a domain-tuned bioinformatics agent confirms the value of specialized tooling for specialized tasks.

{\newtable
\subsection{Common protein analysis tasks}\label{subsec:commontasks}
In order to evaluate the system quantitatively, we created the \emph{Common Protein Analysis Tasks} benchmark, which measures performance on routine workflows and enables controlled comparisons between system configurations. The benchmark contains 50 questions extracted from recurring user usage patterns and divided into five groups of ten, each covering a common task in cheminformatics or protein analysis. Structure analysis covers contacts, hydrogen bonds, solvent exposure, binding pockets, conformational changes, metal coordination, covalent bonds, and construct mutations; molecular visualization covers loading, aligning, modifying, selecting, colouring, and measuring structures in the viewer; 3D reasoning asks the model to select ligand, peptide, or protein placements that are compatible with the shape and chemistry of a binding site or interface; retrieval and selection covers database searches, sequence and residue mappings, structure selection under defined criteria, and cases in which an expected record is absent; and cheminformatics data analysis covers data quality, molecular descriptors, substructure and scaffold trends, selectivity, and activity cliffs. Each question was evaluated with three seeds using deterministic checks.

We compared the full system with three ablations: GPT-5.4-mini instead of GPT-5.4, a configuration without the interactive 3D viewer, and a base configuration with only a Bash shell and no domain tools, skills, or viewer. Molecular visualization was evaluated only for the two configurations with a viewer, while performance across the comparable groups was similar for all configurations (\cref{fig:commontasks}). The clearest effect of the viewer was in 3D reasoning, where accuracy increased from 83.3\% without the viewer to 96.7\% with it, although systems without the viewer could still solve many questions by generating static images for the model to inspect or by constructing ASCII sketches. Future versions could make this group more challenging by requiring more complex visual interpretation and reducing the extent to which answers can be obtained directly from calculations on the atomic coordinates.

\begin{figure}[H]
\centering
\includegraphics[width=\columnwidth]{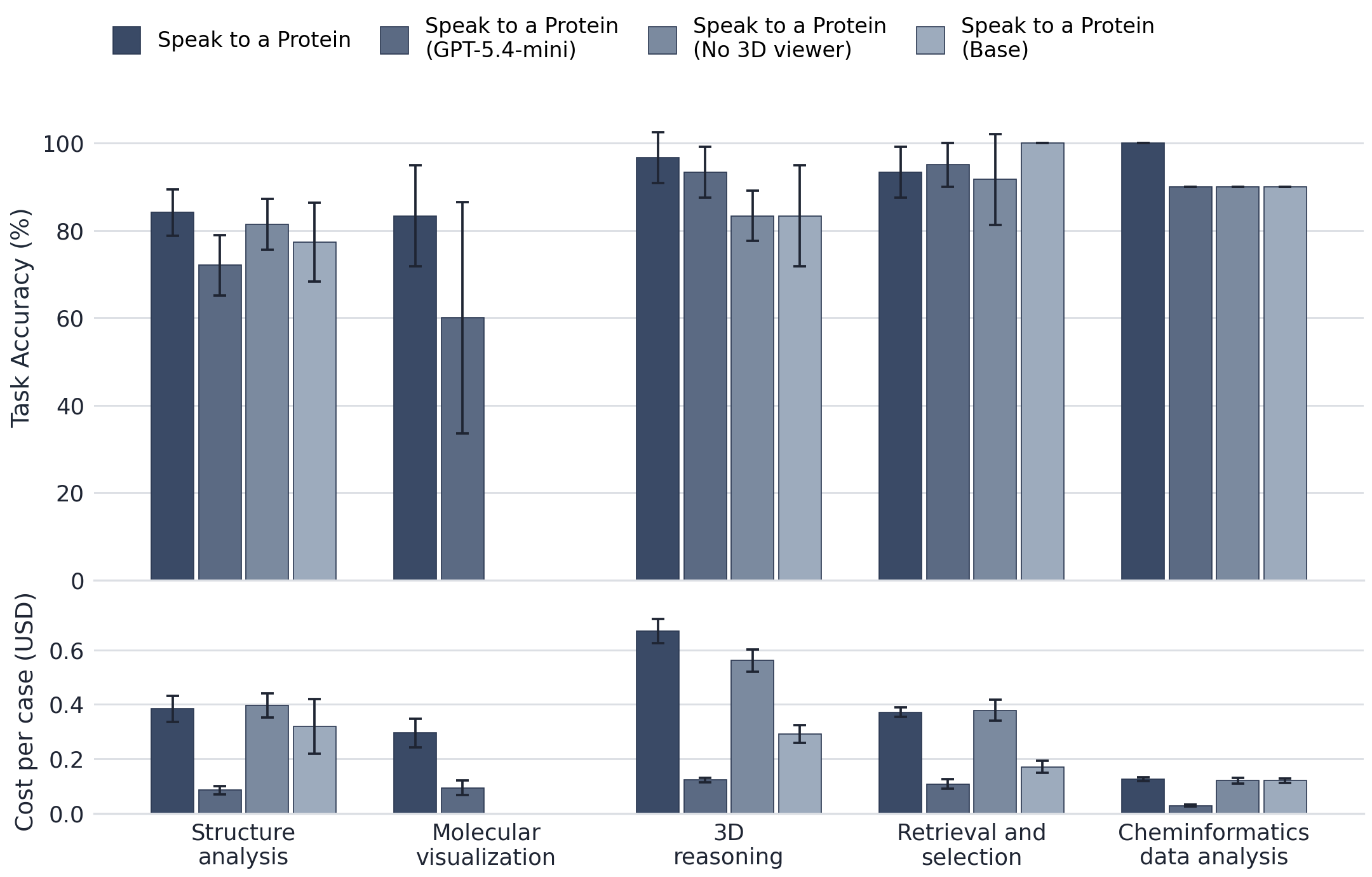}
\caption{Task accuracy and cost on \emph{Common Protein Analysis Tasks}, by group and ablation. Bars are the mean over three seeds and error bars their standard deviation. The visualization group has no bar for the ablations without a viewer, because those tasks cannot be attempted without it.}
\label{fig:commontasks}
\end{figure}
}

\subsection{Speaking about the Dopamine D3 receptor}
The D3R is a G protein-coupled receptor of significant pharmacological interest. Selective ligand design is particularly challenging because of the closely related D2 receptor (D2R), whose orthosteric binding pocket is largely conserved with that of D3R. We illustrate how a user can extract structural and pharmacological conclusions from \emph{Speak to a Protein} through three natural-language prompts, with the agent autonomously coordinating viewer manipulations, codebase and literature search, and bioactivity queries to answer each question. The full session is shown in Video~S1.

\begin{figure}[H]
    \centering
    \begin{subfigure}[b]{0.54\textwidth}
        \centering
        \includegraphics[width=\textwidth]{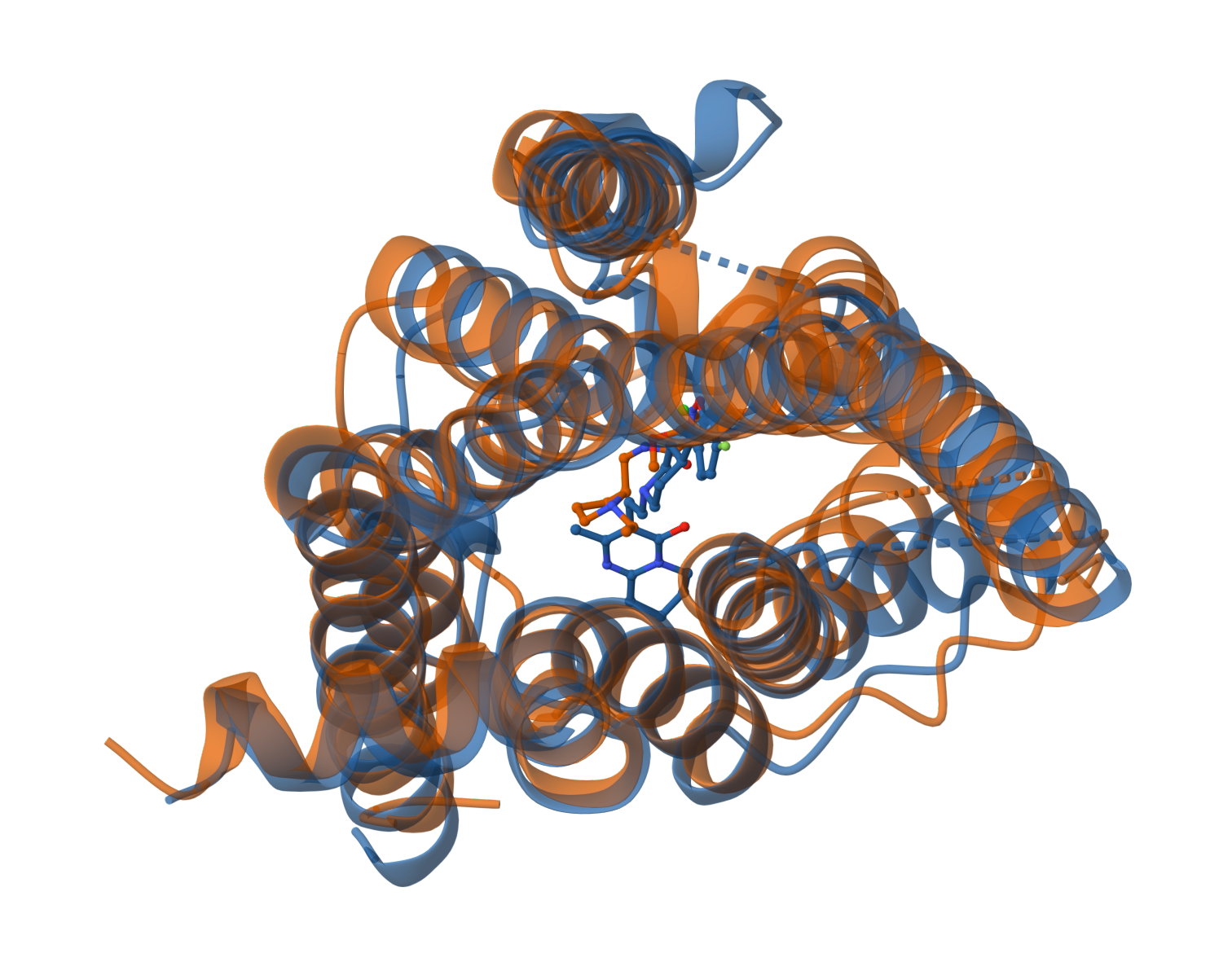}
        \caption{ }
        \label{fig:d3r_align}
    \end{subfigure}
    \hfill
    \begin{subfigure}[b]{0.45\textwidth}
        \centering
        \includegraphics[width=\textwidth]{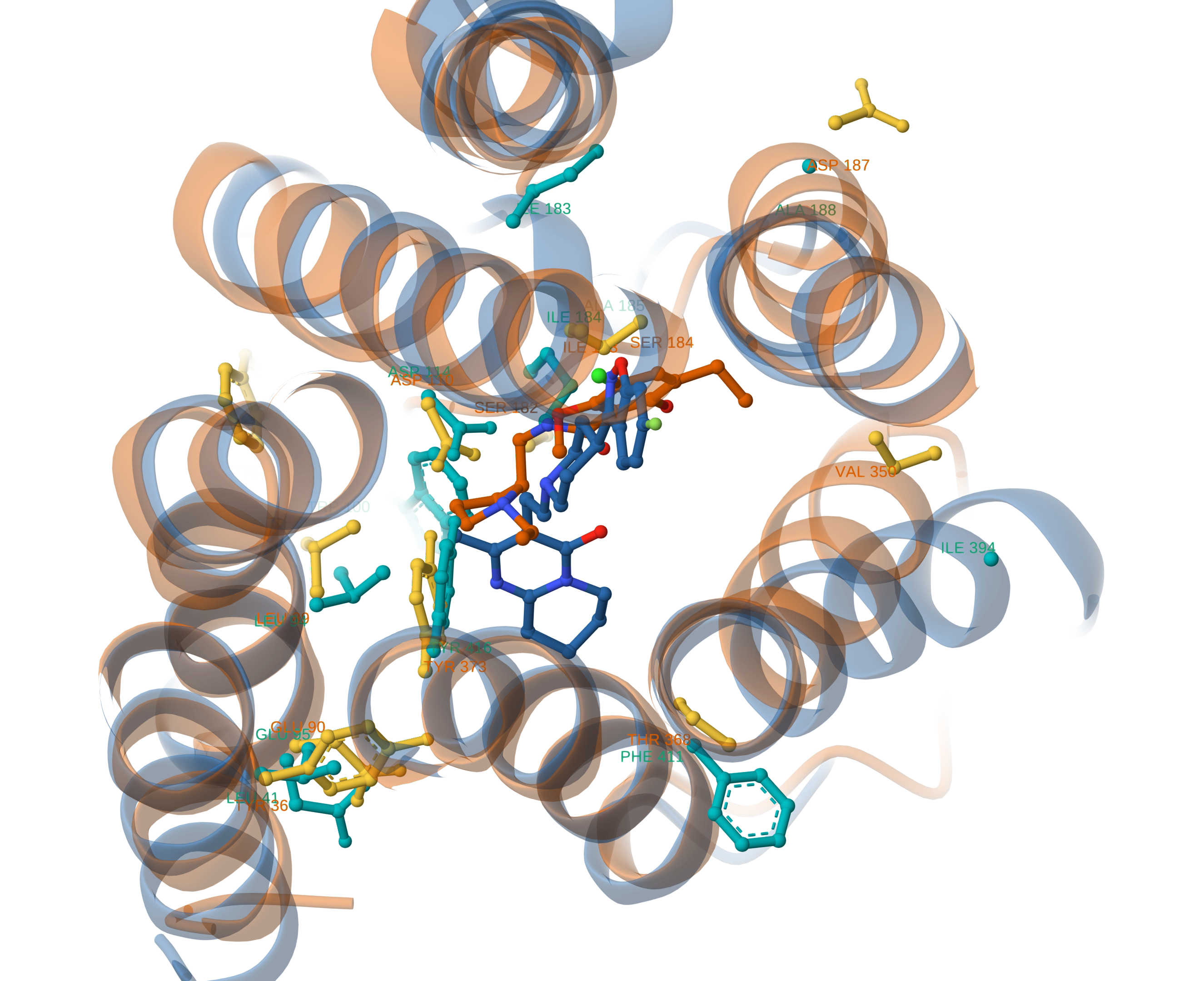}
        \caption{ }
        \label{fig:d3r_residues}
    \end{subfigure}

    \vspace{1em}

    \begin{subfigure}[b]{\textwidth}
        \centering
        \includegraphics[width=\textwidth]{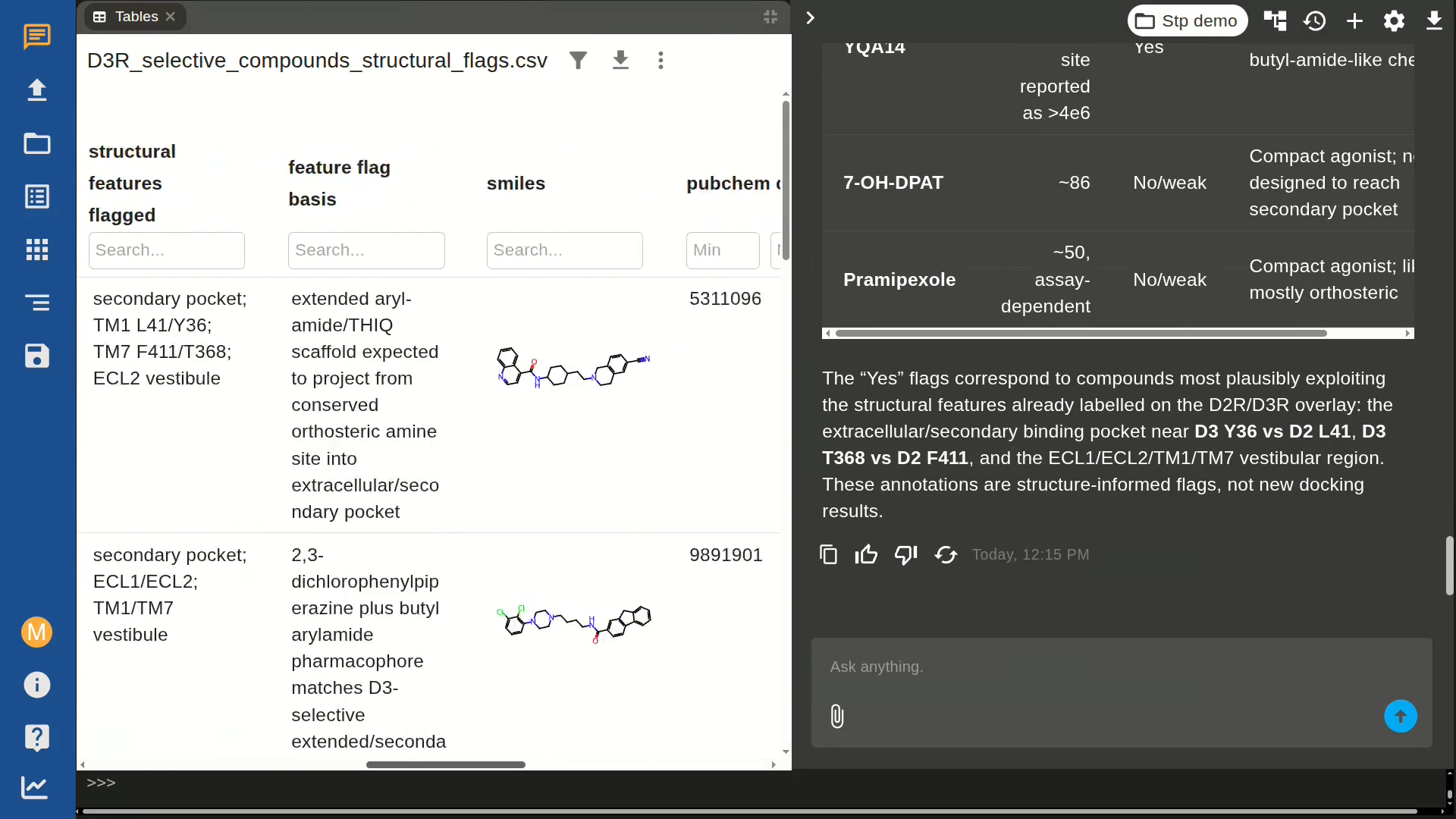}
        \caption{ }
        \label{fig:d3r_final}
    \end{subfigure}

  \caption{\textbf{Structural workflow of the D3R/D2R selectivity case study.} (\subref{fig:d3r_align}) D2R (PDB 6CM4, risperidone-bound, blue) overlaid on D3R (PDB 3PBL, eticlopride-bound, orange) after C\textalpha{} alignment, with the orthosteric ligands as licorice. (\subref{fig:d3r_residues}) Close-up of the orthosteric and adjacent secondary pocket; selectivity-relevant residues are labelled for D2R (cyan) and D3R (yellow), including the conserved orthosteric Asp$^{3.32}$ anchor (D2R Asp114/D3R Asp110). (\subref{fig:d3r_final}) The \emph{Speak to a Protein} interface at the end of the case study, showing the curated CSV of D3R-selective ligands (left) and the chat panel (right).}
    \label{fig:d3r_appendix}
\end{figure}

When studying selectivity between D2R and D3R, a natural starting point is to compare the two receptors directly. A suitable first prompt is therefore \emph{``load representative structures of D2R and D3R and align them,''}. The agent consults the viewer-skill documentation and uses \emph{Codebase Search} over MoleculeKit \citep{Doerr2016} to locate the appropriate alignment routine. It selects two canonical antagonist-bound entries, PDB 6CM4 (D2R bound to risperidone) and 3PBL (D3R bound to eticlopride) \citep{Chien2010,Wang2018}, fetches them from the RCSB, removes the fusion partners introduced for crystallization, and aligns D3R onto D2R on receptor \textalpha-carbons. The aligned structures are loaded into the 3D viewer and rendered, displaying the protein structures and the orthosteric ligands (\cref{fig:d3r_align}).

With the two receptors aligned, the next question is what actually differs between them. A natural follow-up prompt is therefore \emph{``what structural features could be exploited for ligand selectivity between D2 and D3 receptors? Clearly display and label them on the structures.''} The agent issues a \emph{Literature Search} query against the per-protein RAG corpus assembled from UniProt-curated references for both receptors, computes per-residue divergence around the binding pocket using MoleculeKit, and renders the divergent residues on the aligned structures (\cref{fig:d3r_residues}). It identifies the conserved orthosteric anchor (D3R Asp110$^{3.32}$/D2R Asp114$^{3.32}$; superscripts denote Ballesteros-Weinstein numbering\new{, a standard GPCR scheme indexing positions relative to the most conserved residue in each transmembrane helix, assigned index 50 \citep{BALLESTEROS1995366}}) as essential but not selectivity-determining, and singles out four divergent regions: a transmembrane helix 1 (TM1) substitution (D3R Tyr36$^{1.39}$/D2R Leu41$^{1.39}$), an extracellular loop 2 (ECL2) polarity change across several residues (e.g., D3R Ser182/D2R Ile183), a small TM6 volume change (D3R Val350$^{6.56}$/D2R Ile394$^{6.56}$), and a substantial TM7 hydrophobic-to-polar substitution (D3R Thr368$^{7.38}$/D2R Phe411$^{7.38}$). The agent verifies the PDB-to-UniProt residue mapping for both structures, and flags the engineered L119$^{3.41}$W stabilizing mutation in 3PBL to exclude it from the analysis. From these observations it proposes a bitopic ligand strategy: retain the conserved Asp$^{3.32}$ amine interaction while projecting an extended substituent into the divergent extracellular secondary pocket.

To test whether the proposed features are actually exploited in practice, a natural next step is to check the bioactivity data for known selective compounds. A suitable final prompt is therefore \emph{``build a table of known D3R-selective compounds, including their selectivity ratios and SMILES, and load it into the viewer, flagging the ones that engage the structural features identified above.''} The agent draws on the ChEMBL and PubChem skills to query bioactivity records for both UniProt entries, computes median \(K_i\) ratios across published assays, and compiles a curated set of 10 D3R-selective ligands. It cross-references each chemotype against the structural features identified in the previous turn, flagging extended arylamide and arylpiperazine scaffolds (FAUC 365, VK4-116, PG01037, NGB 2904, SB-277011A) as engaging the secondary pocket, and compact agonists (7-OH-DPAT, pramipexole) as orthosteric-only. The compiled CSV is loaded into the viewer's tabular panel alongside the annotated 3D scene (\cref{fig:d3r_final}), producing a single interactive workspace in which structures, selectivity ratios, and chemotypes are coupled to the receptor view.

\new{Because every step of the session is grounded in a database record or a coordinate file, its output can be checked. The three turns make 28 statements with a primary source behind them: which structures were used and in which state, the divergent residues and their Ballesteros-Weinstein positions, and the potency values and scaffolds of the compound table, and all 28 of them are correct. The three turns issued 56 model requests and used 2.8 million input tokens and 34 thousand output tokens, at a cost of \$1.90. Appendix~S1.5 describes the check and reports it by category.}

Beyond the specific assessment of D2/D3 selectivity, the session illustrates the working pattern \emph{Speak to a Protein} is designed to enable. Three coupled natural-language questions produced an integrated analysis that combined viewer manipulations, \emph{Codebase Search} over MoleculeKit, the \emph{Literature Search} RAG, and ChEMBL/PubChem bioactivity queries, carrying a structural hypothesis from aligned receptors to a concrete D3R-selective ligand catalogue.

\subsection{Speaking about the Cyclin-dependent kinase 2}\label{subsec:cdk2}

We illustrate \emph{Speak to a Protein} on a complete drug-discovery workflow centered on CDK2 \citep{Malumbres2009}, a validated yet contested oncology target with three decades of structural and bioactivity data. The case study tests whether the system can take an open-ended question about a target and, from a single user prompt, produce a coherent assessment that integrates structural data, bioactivity, binding-site analysis, and the relevant literature into a shareable report. The prompt issued to the agent is shown below; no further guidance was provided during the session.

\begin{quote}\itshape
``I want to evaluate CDK2 as a drug-discovery target. Work autonomously and explore the problem thoroughly: begin by forming your own research plan, then execute it step by step. Cover the available structural data, the associated bioactivity, the binding site, and the relevant literature, deciding yourself how best to combine these sources. Use the 3D viewer actively: align and inspect the most informative structures, highlight key binding-site residues and ligand interactions, and capture annotated snapshots you can embed in the final document.

Your goal is to synthesize the evidence into a practical drug-discovery assessment. Identify which CDK2 structures and ligand complexes are most informative, what binding-site features matter for inhibitor design, what potency and selectivity trends are supported by data, what the literature says about CDK2 as a therapeutic target, and what next steps medicinal chemists should consider.

The final report should be easy to follow, visual, and self-contained: reference an annotated 3D image wherever you discuss structural features (residues, pockets, ligand interactions), include the relevant plots, and write clear, well-structured prose suitable to share with medicinal chemists.''
\end{quote}

From this single instruction the agent drafted its own research plan and executed it in four stages. It first resolved the canonical human CDK2 entry (UniProt P24941) and retrieved 498 linked PDB structures from RCSB. It then queried ChEMBL for 3{,}394 single-protein CDK2 activity records, together with matched datasets for CDK1, CDK4, CDK6, CDK7, CDK9, and the CDK2/cyclin~E1 and CDK2/cyclin~A2 complexes. From the structural collection it selected representatives spanning apo, cyclin-bound, orthosteric, allosteric, macrocyclic, and ternary-degrader states, loaded them into the 3D viewer, aligned them, and captured annotated snapshots of the cyclin-driven activation transition, the ATP-binding site, and the cyclin~E-dependent interface pocket. Finally, it pulled focused literature on CDK2 biology, cyclin-E dependency, allosteric chemistry, and degrader strategies, and compiled the full analysis into a self-contained LaTeX report reproduced verbatim in Appendix~S1.4.

The agent's main conclusions can be summarized as follows. CDK2 is presented as a structurally well-characterized target, with 489 X-ray and 9 cryo-EM entries at a median resolution of \SI{2.0}{\angstrom} and \new{166} cyclin complexes. Its ATP-binding site is described as broadly addressable, with a recurring contact set (Ile10, Ala31, Lys33, the Phe80--Leu83 hinge, Gln131--Asn132, Leu134, Asp145) that explains why subnanomolar potency is so often achieved. Selectivity is identified as the main challenge: across compounds with matched assays, the median CDK2 advantage over CDK1 is only 0.32 pChEMBL units, just 15.3\% reach a one-log margin, and the overlap against CDK9 is similar. To address this, the report points to two structural alternatives to the ATP site, namely the ANS allosteric site \citep{betzi2011allosteric} and the more recent cyclin~E-dependent interface pocket exemplified by PDB 8VQ3, and links them to a biomarker-led therapeutic case in CCNE1-amplified disease, cyclin-E-high tumors, and settings of CDK4/6-inhibitor resistance, supported by recent work on replication-stress biology \citep{brown2023forkrepair} and CDK2/cyclin~E heterobifunctional degraders \citep{kwiatkowski2025degrader}. The agent's overall recommendation is that CDK2 is a compelling target when biology, modality, and a selectivity strategy are aligned, but less attractive as a broad oncology kinase project.

\begin{figure}[H]
\centering
\begin{subfigure}[b]{0.38\columnwidth}
    \centering
    \includegraphics[height=2.6in,keepaspectratio]{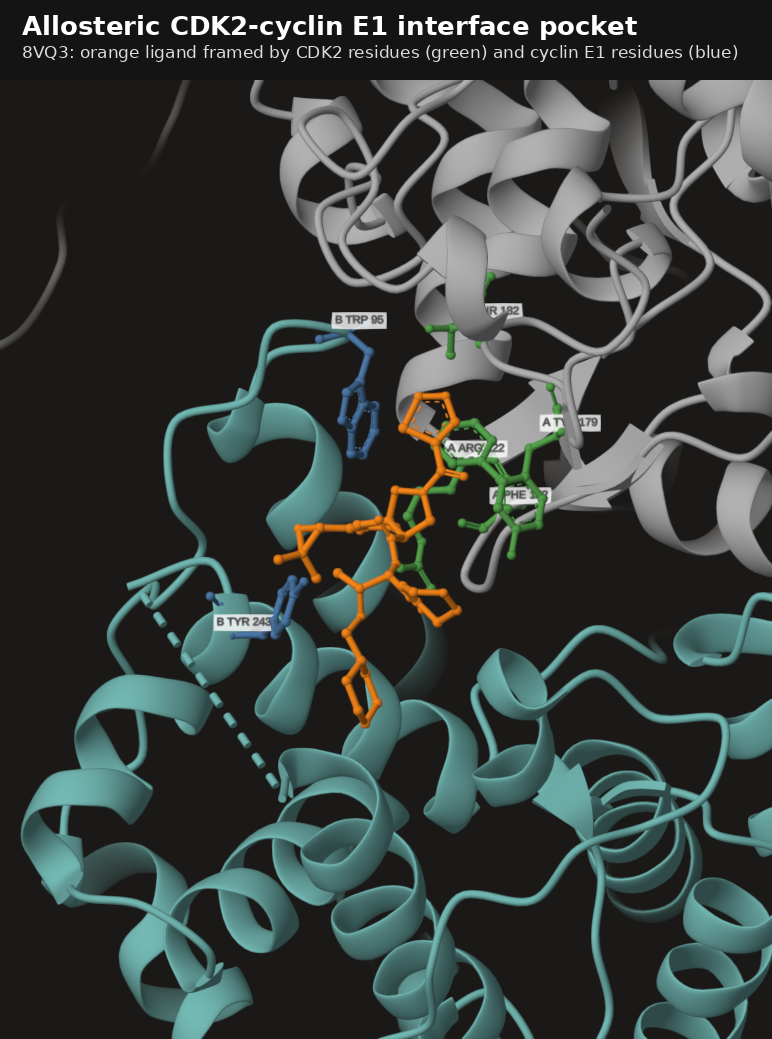}
    \phantomcaption\label{fig:cdk2_allosteric}
\end{subfigure}\hfill
\begin{subfigure}[b]{0.6\columnwidth}
    \centering
    \includegraphics[height=2.6in,keepaspectratio]{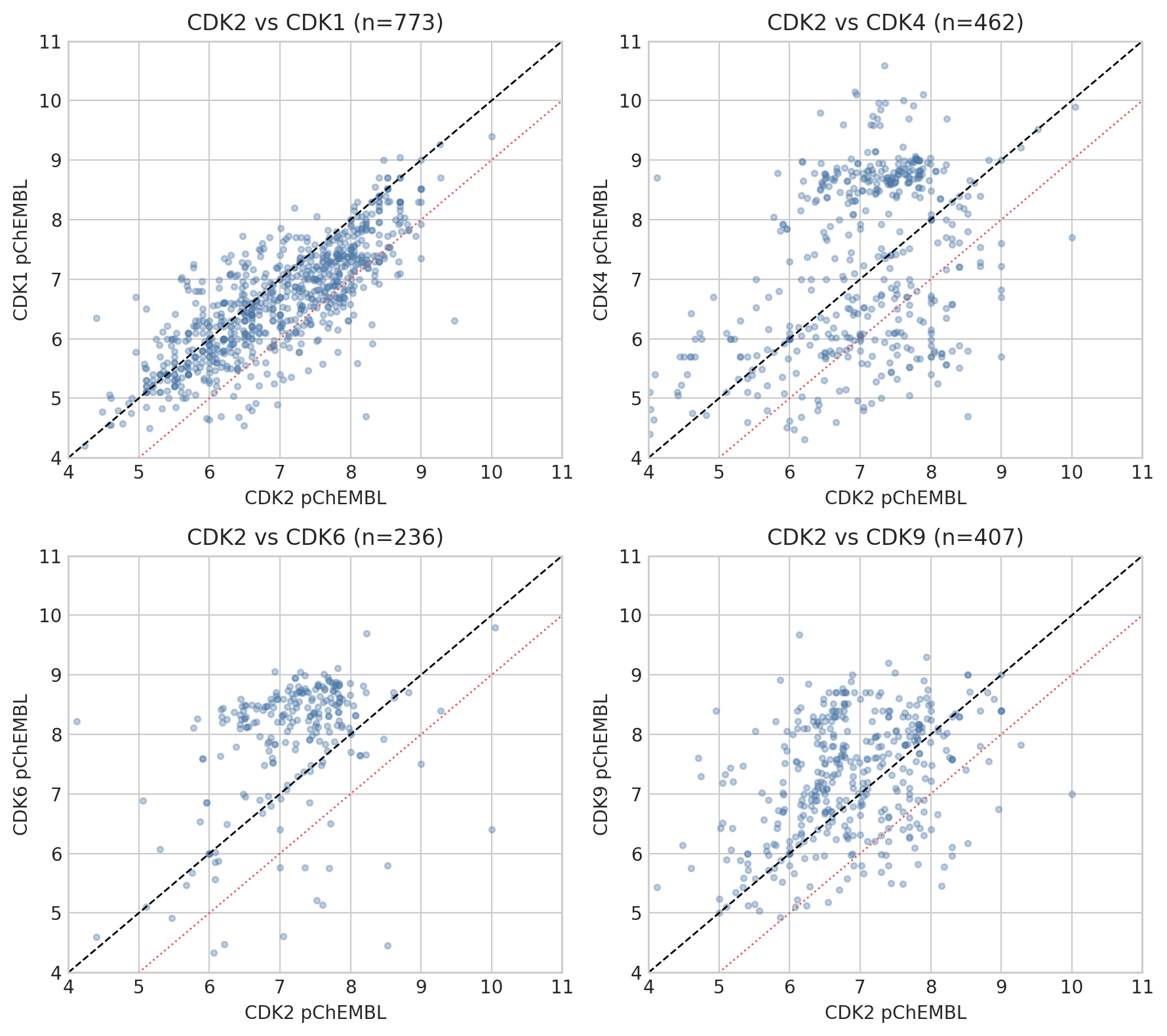}
    \phantomcaption\label{fig:cdk2_selectivity}
\end{subfigure}
\caption{Figures the agent generated and embedded in its own report. \textbf{(a)} Annotated 3D-viewer snapshot of the cyclin~E-dependent allosteric interface pocket (PDB 8VQ3), jointly framed by CDK2 (green) and cyclin~E1 (blue) residues, captured by the agent during the analysis. \textbf{(b)} Matched-compound CDK2 selectivity relative to CDK1, CDK4, CDK6, CDK7, and CDK9 from filtered ChEMBL bioactivity; the dashed line is parity, the dotted red line marks a one-log margin in favor of CDK2.}
\label{fig:cdk2_analysis}
\end{figure}

Two figures from the agent's own report are reproduced in \cref{fig:cdk2_analysis}: an annotated 3D-viewer snapshot of the cyclin~E-dependent allosteric interface pocket, and a matched-compound selectivity panel for CDK2 against five related kinases. Together they show how the agent combined direct structural interpretation with quantitative analysis. The remaining figures, tables, and supporting data accompany the full report in the appendix.

\new{The report makes 108 statements with a primary source behind them, covering the structure counts, the two tables, the distances measured from the coordinates, the bioactivity statistics and the references, and of these 107 are correct. The one exception is the number of cyclin complexes, reported as 374 in the original generated report, where the correct number is 166, because the script the agent wrote counted every entry whose chain description contains the word \emph{cyclin}, and that word also appears in the name of the kinase itself. The run issued 91 model requests and used 7.5 million input tokens and 60 thousand output tokens, at a cost of \$4.00. Appendix~S1.5 describes the check and reports it by category.}

Beyond the specific assessment of CDK2, the run illustrates the working pattern \emph{Speak to a Protein} is designed to enable. A single open-ended question produced a connected workflow that combined autonomous planning, multimodal data integration, agentic use of the 3D viewer, and a publication-ready report within one session.

\section{Conclusions}\label{sec:discussion}

We presented \emph{Speak to a Protein}, a multimodal agentic co-scientist that turns protein analysis into an interactive dialogue grounded in language, code execution, and a live 3D viewer. The system retrieves and integrates information across literature, structural repositories, and bioactivity databases, drives the viewer with natural language, executes analysis code in a sandboxed environment, and uses agentic 3D vision to interpret the rendered scene. On BixBench-Verified-50, a benchmark of agentic bioinformatics and data-analysis tasks, the system reaches 78.5\% accuracy, on par with specialized bioinformatics agents and ahead of strong general-purpose coding agents. The two case studies illustrate the intended working pattern. The D3R interaction shows incremental, conversational use of the viewer, tabular data, codebase search, literature retrieval and database querying. The CDK2 study shows the system carrying a single open-ended drug-discovery question to a coherent target assessment, with structural snapshots, selectivity analysis, and a self-contained report produced in one session. Taken together, these results suggest that pairing a competent agent harness with structural biology tools meaningfully lowers the time and skill barrier between a scientific question and structured, reusable evidence, compressing workflows that traditionally take hours of manual data gathering and analysis into a single interactive session.

\new{One limitation of the system is that language models hallucinate, and this is hard to measure and hard to control. In our experience stronger models hallucinate less, and since the harness is not tied to a particular model, we expect this to improve as models improve.}

\section*{Data and Software Availability}

\emph{Speak to a Protein} is freely accessible at \url{https://open.playmolecule.org}, where reviewers and readers can reproduce the interactive sessions described in this work. A commercial version is available at \url{https://playmolecule.ai}.

Code to reproduce the results in this paper is available at \url{https://github.com/Acellera/speak-to-a-protein}. The \emph{BixBench} benchmark~\citep{mitchener2025bixbench} and the \emph{BixBench-Verified-50} subset~\citep{phylo2026bixbenchverified} are openly available from their respective sources. The structures and bioactivity records used in the D3R and CDK2 case studies are identified by their accession codes in the main text and Supporting Information. \new{The code that verifies the claims of Appendix~S1.5 and the traces of the two case-study sessions are released in the same repository.}

\section*{Conflict of Interest}

The authors declare the following competing financial interest(s): C.N., M.T.F., P.T., S.D., and G.D.F. have a potential conflict of interest due to direct interests in Acellera (see affiliations).

\begin{acknowledgement}
The project PID2020-116564GB-I00 has been funded by MCIN/AEI/10.13039/501100011033 (G.D.F.). This project has received funding from the Torres-Quevedo Programme from the Spanish National Agency for Research (PTQ2021-011670/AEI/10.13039/501100011033) (M.T.F.). With the support of the Industrial Doctorates Plan of the Secretariat of Universities and Research of the Department of Economy and Knowledge of the Generalitat of Catalonia (C.N.).
\end{acknowledgement}

\begin{suppinfo}
Table of the fields returned by the literature search tool; user and token usage of the public deployment between October 2025 and July 2026; complete report generated by the agent for the CDK2 case study; verification of the statements of the two case studies against their primary sources (PDF).

Video S1, recording of the D3R/D2R selectivity session (MP4).
\end{suppinfo}

\bibliography{references}

\end{document}


\newpage
\section{Appendix}
\subsection{Tables}
\begin{table}[!htbp]
\caption{Fields returned by the \new{literature search tool}. Global fields describe the query context; results contain one or more retrieved citations with text passages.}
\label{tab:literature-tool}
\begin{center}
\begin{tabular}{p{3.7cm}p{10cm}}
\toprule
\multicolumn{1}{c}{\bf Field} & \multicolumn{1}{c}{\bf Description / Example Value} \\
\midrule
\texttt{fasta\_sequences[]} & One or more FASTA sequences associated with the protein input. \\
\addlinespace
\texttt{pdb\_ids[]} & PDB identifiers discovered for the input (via cross-references or sequence search). \emph{Example:} ["1H1R", "3DDQ", "7RWE"] \\
\addlinespace
\texttt{uniprot\_ids[]} & UniProt accessions associated with the input. \emph{Example:} ["P24941"] \\
\addlinespace
\texttt{results[].pmcid} & PubMed Central ID for a retrieved article. \emph{Example:} "PMC5291709" \\
\addlinespace
\texttt{results[].doi} & Digital Object Identifier for the article. \emph{Example:} "10.7554/eLife.20818" \\
\addlinespace
\texttt{results[].content} & Retrieved text passage from the article body (cleaned paragraphs). \emph{Example:} "ATP-binding sites involve Walker-A and -B motifs and a trigger residue that regulates hydrolysis." \\
\addlinespace
\texttt{error} & Error message if something went wrong; empty/absent otherwise. \emph{Example:} "Failed to load existing RAG database: …" \\
\bottomrule
\end{tabular}
\end{center}
\end{table}

\clearpage
\subsection{Usage Metrics} \label{sec:appendix_usage}

\begin{figure}[htbp]
    \centering
    \begin{subfigure}[b]{1\textwidth}
        \centering
        \includegraphics[width=\textwidth]{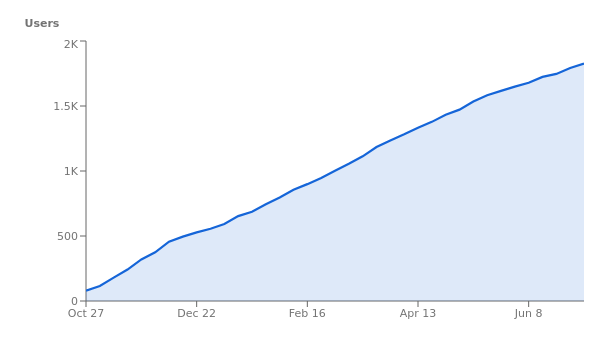}
        \caption{Aggregated number of unique users engaging with the AI Chat.}
        \label{fig:usage_users}
    \end{subfigure}

    \vspace{1em} 

    \begin{subfigure}[b]{1\textwidth}
        \centering
        \includegraphics[width=\textwidth]{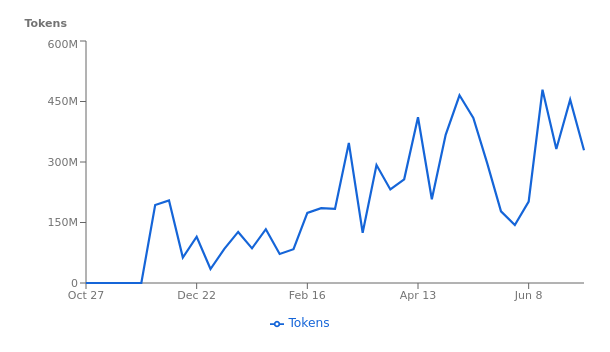}
        \caption{Weekly token consumption across AI Chat sessions.}
        \label{fig:usage_tokens}
    \end{subfigure}

    \caption{\textbf{Platform Usage Statistics.} Usage metrics from the public release of \emph{Speak to a Protein}, demonstrating steady user adoption and a substantial increase in weekly token consumption from late 2025 \new{through mid-2026}.}
    \label{fig:usage_metrics}
\end{figure}

To demonstrate the practical utility and robustness of the system, we analyzed usage statistics from the public deployment of \emph{Speak to a Protein} between October 2025 and \new{July 2026}. These metrics reflect organic adoption by the scientific community. As shown in \cref{fig:usage_metrics}, the platform has seen consistent growth in its user base\new{, reaching 1,921 users of the AI chat}.  Simultaneously, the volume of interaction has accelerated. \cref{fig:usage_tokens} displays weekly token consumption, which increased steadily over the monitoring period, peaking at \new{479 million tokens in June 2026}.

\clearpage
\subsection{Structural workflow of the D3R/D2R selectivity case study.}\label{sec:appendix_d3r}

\noindent\textbf{Video~S1.} Full \emph{Speak to a Protein} session for the D3R/D2R selectivity case study, illustrating the three-prompt workflow from structure loading and alignment, through residue-level selectivity analysis, to the assembly of the D3R-selective compound table.

\clearpage
\subsection{Auto-Generated CDK2 Analysis Report}\label{sec:appendix_cdk2_report}

The following pages reproduce the complete report generated by \emph{Speak to a Protein} for the CDK2 case study, exactly as written by the agent from the single prompt shown in Section~4.4 of the main paper. No content has been altered.

\includepdf[pages=-,pagecommand={\thispagestyle{plain}}]{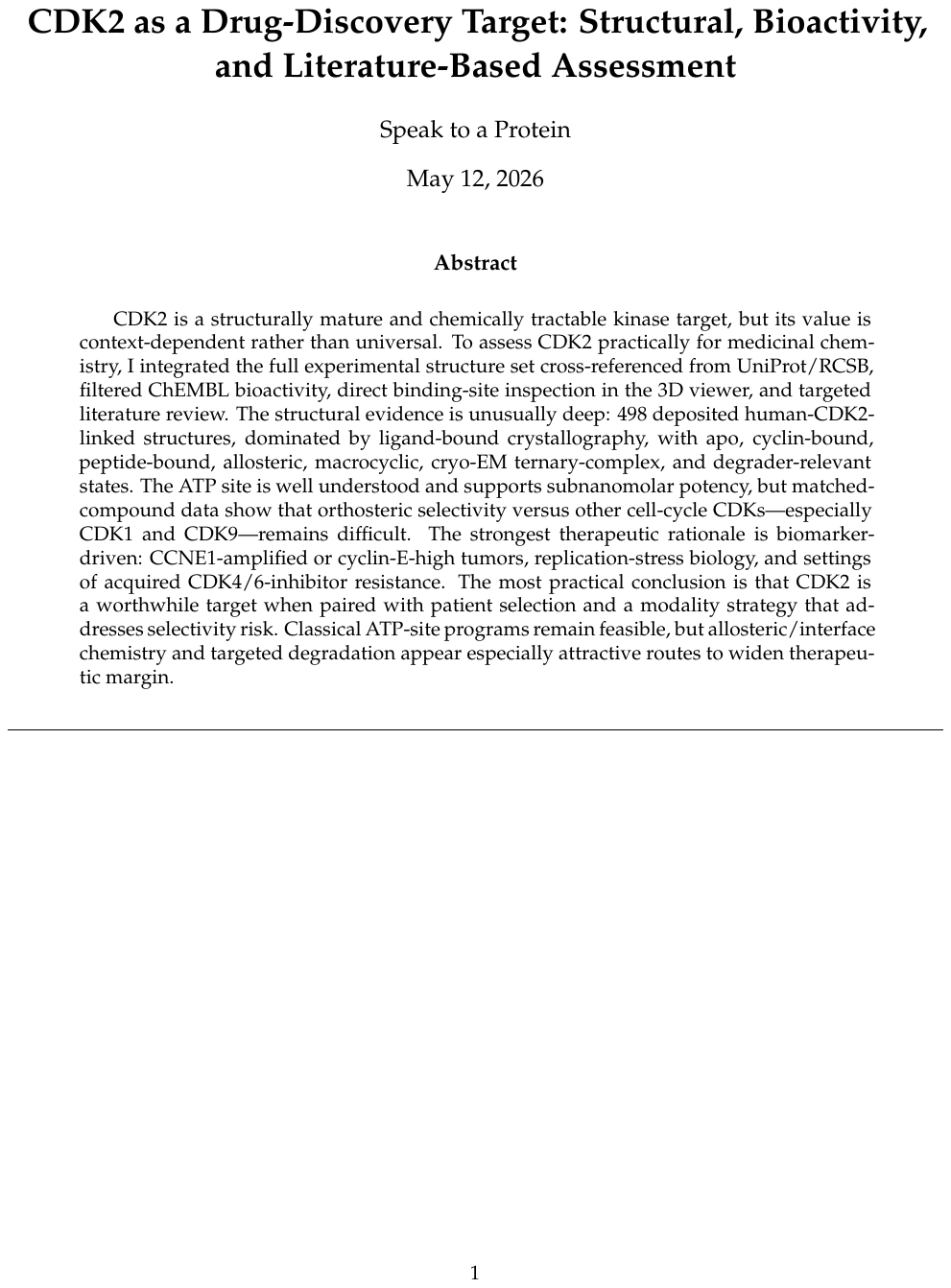}

\clearpage
{\newtable
\subsection{Reliability of the agent's case studies}\label{sec:appendix_audit}

Every verifiable claim of the two case studies was checked against its primary source: 75 claim entries that resolve into the 136 individual statements of \cref{tab:audit_full}, each one a number, a name or an assignment. Judgements were left out, because they have no ground truth. Where an independent route existed we took it rather than reuse the agent's own tools: coordinates were read with a parser written for this check, the D3R and D2R positions were matched through the GPCRdb generic numbering instead of the alignment the session computed, and every reference was resolved through Crossref and Europe PMC. Both databases moved after the sessions ran, so each query is anchored to the state the agent saw, version 275 of UniProt entry P24941, which lists the 498 structures the agent retrieved where the entry today lists 512, and release 36 of ChEMBL, which the web service still served on both session dates. Of the 136 statements, 135 are correct. The one error is the number of cyclin complexes, given as at least 374 where the correct number is 166: the session counted every entry whose chain description contains the word \emph{cyclin}, and that word also appears in \emph{Cyclin-dependent kinase 2}, the name of the kinase itself, so 208 of the 374 entries carry no cyclin chain. The claim files, the scripts and the evidence retrieved for each claim are released with the paper.

\begin{table}[!htbp]
\caption{Verifiable statements of the two case studies, checked against their primary sources at the database state each session saw.}
\label{tab:audit_full}
\begin{center}
\small
\setlength{\tabcolsep}{4pt}
\begin{tabular}{llrr}
\toprule
\textbf{Case study} & \textbf{Category} & \textbf{Statements} & \textbf{Reproduced} \\
\midrule
CDK2 & Structure counts and resolution statistics & 8  & 7 \\
CDK2 & Identity, state and resolution of a named entry & 9  & 9 \\
CDK2 & Distances and contacts from coordinates & 34 & 34 \\
CDK2 & Counts and statistics over the activity export & 8  & 8 \\
CDK2 & Potency of a named compound & 35 & 35 \\
CDK2 & Compound and ligand identity & 2  & 2 \\
CDK2 & References resolved and supporting their statement & 12 & 12 \\
D3R  & Identity and state of a named entry & 3  & 3 \\
D3R  & Residue assignments and generic numbering & 7  & 7 \\
D3R  & Potency values and selectivity ratios & 11 & 11 \\
D3R  & Compound scaffold assignments & 7  & 7 \\
\midrule
\textbf{Total} & & \textbf{136} & \textbf{135} \\
\bottomrule
\end{tabular}
\end{center}
\end{table}
}